\documentstyle[aps,preprint,epsf,rotate]{revtex} 
\tighten

\begin{document}
\draft
\widetext
\preprint{hep-ph/9610248} 

\title{Supersymmetric Inflation and Large-scale Structure
\footnote{Work done in collaboration with J.A. Adams \& G.G. Ross [1]}}
\author{Subir Sarkar \vspace{2mm}}
\address{Theoretical Physics, University of Oxford, \\
         1 Keble Road, Oxford OX1 3NP, U.K.}
\date{\tt sarkar@thphys.ox.ac.uk}
\maketitle

\begin{abstract} 
In effective supergravity theories following from the superstring, a
modulus field can quite naturally set the neccessary initial
conditions for successful cosmological inflation to be driven by a
hidden sector scalar field. The leading term in the scalar potential
is {\em cubic} hence the spectrum of scalar density perturbations
neccessarily deviates from scale-invariance, while the generation of
gravitational waves is negligible. The growth of large-scale structure
is then consistent with observational data assuming a critical density
cold dark matter universe, with no need for a component of hot dark
matter. The model can be tested thorough measurements of cosmic
microwave background anisotropy on small angular scales.
\end{abstract}

\pacs{\center{\sf Invited talk at the Inauguration Conference of the
Asia-Pacific Center for Theoretical Physics, Seoul (June 4-10, 1996)}}

\small

\section{Introduction}

It is well known that a sufficiently long period of accelerated,
non-adiabatic expansion in the early universe, driven by the false
vacuum energy of a scalar field, successfully solves the horizon and
flatness problems of the standard Big Bang model, as well as the
cosmological monopole problem of grand unified theories \cite{guth}.
In the `new' inflationary model \cite{book}, a single bubble of the
true vacuum expands sufficiently in the vacuum energy dominated De
Sitter epoch, so as to contain the entire universe visible today and
drive it to the critical density; the vaccum energy is then converted
to radiation, `reheating' the universe and starting off the standard
Friedmann-Robertson-Walker evolution. Furthermore, the density
perturbations generated by quantum fluctuations of the scalar field
driving inflation have a (nearly) scale-invariant spectrum, as is
required by observations \cite{book}. However it was observed over a
decade ago that in order to respect the observational limit on the
perturbation amplitude deduced from the isotropy of the 2.73\,K cosmic
microwave background (CMB), the scalar potential has to be extremely
flat and protected against radiative corrections. It became evident
that the only plausible candidates for the `inflaton' are gauge
singlet fields in supersymmetric theories \cite{susyinflrev}, which
were recognized already as being the most probable extension of
physics beyond the Standard
$SU(3)_{c}{\otimes}SU(2)_{L}{\otimes}U(1)_{Y}$ Model \cite{sugrarev}.

However all such models \cite{susyinflrev} were found to be plagued
with various phenomenological problems, in particular the production
during reheating of massive unstable particles such as gravitinos
whose late decays can disrupt the standard cosmology \cite{gravprob}
as well as the excitation during inflation of weakly coupled scalar
fields associated with supersymmetry breaking, which too release their
energy rather late generating an unacceptable amount of entropy
\cite{polonyiprob}. (The latter problem is particularly acute for the
flat directions (or moduli) of string theories
\cite{moduliprob,success}.) It was also established that thermal
effects in the early universe cannot localize the inflaton field at
the origin as is required to ensure a sufficiently long period of
inflation \cite{susyinflrev}. However given {\em random} initial
conditions, as is appropriate for a weakly coupled field, successful
inflation was shown to be possible if the inflaton has its global
minimum at the origin and evolves towards it from an initial vacuum
expectation value (vev) beyond the Planck scale. Such a `chaotic'
inflation model \cite{book} accomodates a wide variety of potentials
(albeit with arbitrary fine tuning) hence attention drifted away from
the specific problems encountered by supersymmetric inflationary
models.

Subsequently, precision accelerator data \cite{lep} have confirmed
that the most likely solution to the `hierarchy' problem posed by a
{\em fundamental} Higgs boson in the Standard Model is indeed (broken)
supersymmetry just above the electroweak scale. Moreover, this enables
successful unification of the strong and electroweak gauge couplings
at a scale of $\approx2\times10^{16}$\,GeV as well as providing an
elegant mechanism for electroweak symmetry breaking and an
understanding of the pattern of fermion masses \cite{susyunif}. The
superpartners of the known particles should have masses no higher than
a few TeV so can be directly created at the forthcoming {\sl LHC} or
possibly even at {\sl LEP 2}. The lightest supersymmetric particle is
typically the neutralino, a neutral weakly interacting mixture of the
superpartners of the gauge and Higgs bosons. It naturally has a relic
abundance of order the critical density \cite{susydm} and is therefore
an excellent candidate for the cold dark matter (CDM) \cite{cdm} which
is required in all viable models of large-scale structure formation
\cite{structure}. This provides strong motivation to reexamine the
problems connected with inflation in supersymmetric theories,
specifically $N=1$ supergravity, the phenomenologically successful
effective field theory below the Planck scale \cite{sugrarev}. We
focus on models where supersymmetry breaking occurs in a `hidden'
sector and is communicated to the visible sector through gravitational
interactions.

Meanwhile on the cosmological front, the discovery by {\sl COBE} of
temperature fluctuations in the CMB on angular scales larger than the
causal horizon at (re)combination has provided strong indirect support
for inflation. The power spectrum of the fluctuations is consistent
with a scale-invariant primordial perturbation, and the statistics
with a random Gaussian field, both as predicted by inflation
\cite{cobe}. Thus the spectrum of scalar density perturbations can be
normalized directly to {\sl COBE} (taking into account any
gravitational wave component which would also contribute to the CMB
anisotropy). The primordial spectrum is modified on scales smaller
than the horizon size at matter-radiation equality by a `transfer
function' characteristic of the matter content of the universe
\cite{structure}. The power spectrum inferred from observations of the
clustering and motions of galaxies \cite{efstathiou} can then be
compared with the theory. Of particular interest is whether the
problem of excessive small-scale power in a CDM universe (assuming
{\em scale-invariant} primordial fluctuations) \cite{white} can be
resolved by the spectral `tilt' expected from supersymmetric inflation
\cite{success,eps}, rather than by invoking a component of hot dark
matter. Ongoing and future observations of the CMB anisotropy on small
angular scales \cite{cmbrev} will provide an independent test of this
possibility.

\section{Requirements of the Inflationary Potential}

The main theoretical problem in constructing an inflationary model
based on supergravity is that the large cosmological constant during
inflation breaks global supersymmetry, giving all scalar fields a
`soft' mass of order the Hubble parameter \cite{susybreak}. In the
simplest models, the inflaton potential thus acquires a curvature too
large to allow inflation to proceed for the required number of
e-folds. The problem is characteristic of the scalar potential (along
a F-flat direction) of a singlet field in the hidden sector having
minimal kinetic terms, hence various solutions have been proposed
which modify one or the other of these assumptions, e.g. introduction
of non-minimal kinetic terms, or specific interactions of the inflaton
with gauge fields, or identification of the inflaton as a D- (rather
than F-) flat direction or even as a modulus field
\cite{modelrev}. Here, I would like to discuss a new mechanism leading
to successful inflation in the low-energy effective supergravity
theory following from the superstring \cite{paper}. The interesting
observation is that in a wide class of such theories, the equations of
motion have an infra-red fixed point at which successful inflation can
occur, even for {\em minimal} kinetic terms, along a F-flat direction.

First, let us briefly review the necessary ingredients for successful
inflation with a scalar potential $V(\phi)$. Essentially all model
generating an exponential increase of the cosmological scale-factor
$a$ satisfy the `slow-roll' conditions \cite{inflrev}
\begin{equation}
 \dot{\phi} \simeq -\frac{V'}{3H}\ , \quad
 \epsilon \equiv \frac{M^2}{2} \left(\frac{V'}{V}\right)^2 \ll 1\ ,
  \qquad
 |\eta| \equiv \left|M^2 \frac{V^{''}}{V}\right| \ll 1\ ,
\label{slowroll}
\end{equation}
where $H\simeq\sqrt{V/3M^2}$ is the Hubble parameter during inflation,
and the normalized Planck mass $M\equiv\,M_{\rm
Pl}/\sqrt{8\pi}\simeq2.44\times10^{18}$ GeV. Inflation ends
(i.e. $\ddot{a}$ drops through zero) when $\epsilon,
|\eta|\simeq1$. The spectrum of adiabatic scalar perturbations is
\cite{inflrev}
\begin{equation}
 \delta^2_{\rm H} (k) = \frac{1}{150 \pi^2} \frac{V_{\star}}{M^{4}} 
                      \frac{1}{\epsilon_{\star}}\ ,  
\label{deltah}
\end{equation}
where $\star$ denotes the epoch at which a scale of wavenumber $k$
crosses the `horizon' $H^{-1}$ (more correctly, Hubble radius) during
inflation, i.e. when $aH=k$. The CMB anisotropy measured by {\sl COBE}
\cite{cobe} allows a determination of the fluctuation amplitude at
the scale, $k_{\sl COBE}^{-1}\sim\,H_0^{-1}\simeq\,3000\,h^{-1}$\,Mpc,
corresponding roughly to the size of the presently observable
universe, where $h\equiv\,H_0/100$\,km\,sec$^{-1}$\,Mpc$^{-1}$ is the
present Hubble parameter. The number of e-folds before the end of
inflation when this scale crosses the Hubble radius is
\begin{eqnarray}
 N_{\sl COBE} \equiv N_\star (k_{\sl COBE}) & \simeq 51 
  & + \ln\left(\frac{k_{\sl COBE}^{-1}}{3000h^{-1}\,{\rm Mpc}}\right) 
    + \ln\left(\frac{V_\star}{3\times10^{14}\,{\rm GeV}}\right) 
    + \ln\left(\frac{V_\star}{V_{\rm end}}\right) \nonumber \\
 && - \frac{1}{3}\ln\left(\frac{V_{\rm end}}{3\times10^{14}\,{\rm GeV}}\right) 
    + \frac{1}{3}\ln\left(\frac{T_{\rm reheat}}{10^5\,{\rm GeV}}\right),
\label{Nstar} 
\end{eqnarray}
where we have indicated the numerical values anticipated for the
various energy scales in our model. (Note that $N_{\sl COBE}$ is
smaller than the usually quoted \cite{inflrev} value of 62 because the
reheat temperature {\em must} be low enough to suppress the production
of unstable gravitinos which can disrupt primordial nucleosynthesis
\cite{gravprob}.) The {\sl COBE} observations sample CMB multipoles
upto $\sim20$, where the $l^{\rm th}$ multipole probes scales around
$k^{-1}\sim6000h^{-1}l^{-1}\,{\rm Mpc}$. The low multipoles, in
particular the quadrupole, are entirely due to the Sachs-Wolfe effect
on super-horizon scales ($k^{-1}>k_{{\rm
dec}}^{-1}\simeq\,180h^{-1}$\,Mpc) at CMB decoupling and thus a direct
measure of the primordial perturbations. However the high multipoles
are (increasingly) sensitive to the composition of the dark matter
which determines how the primordial spectrum is modified through the
growth of the perturbations on scales smaller than the horizon at the
epoch of matter-radiation equality, i.e. for $k^{-1}<k_{\rm
eq}^{-1}\simeq\,80h^{-1}$\,Mpc. Thus the normalization of the spectrum
(\ref{deltah}) to the {\sl COBE} data is sensitive to its $k$
dependence and also on whether there is a contribution from
gravitational waves to the CMB anisotropy. The 4-year data is fitted
by a scale-free spectrum, $\delta^2_{{\rm H}}\sim\,k^{n-1}$,
$n=1.2\pm0.3$, with $Q_{\rm rms}=15.3^{+3.8}_{-2.8}\,\mu$K
\cite{cobe}. For a scale-invariant ($n=1$) spectrum, $Q_{\rm
rms}=18\pm1.6\,\mu$K, so assuming that there are no gravitational
waves, the amplitude for a $\Omega=1$ CDM universe is $\delta_{{\rm
H}}=(1.94\pm0.14)\times10^{-5}$ \cite{cobenorm}. Using
eq.(\ref{deltah}), the vacuum energy at this epoch is then given by
\begin{equation}
 V_{\sl COBE} \simeq (6.7 \times 10^{16}\,{\rm GeV})^4\ \epsilon_{\sl COBE}\ ,
\label{scale}
\end{equation}
showing that the inflationary scale is far below the Planck scale
\cite{inflrev}. A similar limit obtains, viz. $V_{\sl
COBE}\lesssim(4.9\times10^{16}\,{\rm GeV})^4$, if the observed
anisotropy is instead ascribed entirely to gravitational waves, the
amplitude of which, in ratio to the scalar perturbations, is just
\cite{inflrev}
\begin{equation}
 r = 12.4\ \epsilon\ .
\label{grav}
\end{equation}
Thus it is legitimate to study inflation in the context of an
effective field theory. The potential then has the generic form
\begin{equation}
 V \sim \Lambda^4 \left[1 + c_n (\phi/M)^n\right]\ .
\end{equation}
In the usual model of chaotic inflation, one has $\phi/M\gg1$ so the
first term on the rhs is negligible and $\epsilon$ and $\eta$ are
small because they are proportional to $(\phi/M)^{-2}$. Alternatively
$\phi$ may start with a vev much smaller than the Planck scale during
inflation, in which case the smallness of $V'$ and $V''$, and hence
$\epsilon$ and $\eta$, results from the relative smallness of the
second term on the rhs.

To take into account both cases, let us expand the (slowly varying)
potential about the value $\phi^{*}$ in inflaton field space at which
the observed density fluctuations are produced. Writing
$\phi=\tilde{\phi}+\phi^{*}$ (in units of $M$) we have
\begin{equation}
 V (\phi) = \Lambda^4 \left[1 + c_1 \tilde{\phi}
                             + c_2 \tilde{\phi}^2
                             + c_3 \tilde{\phi}^3 
                             + c_4 \tilde{\phi}^4
                             + \dots \right]\ .  
\label{expand}
\end{equation}
Here we have factored out the overall scale of inflation $\Lambda$,
which we have seen must be small relative to the Planck scale $M$. The
constraints on the parameters in the potential following from the
slow-roll conditions (\ref{slowroll}) are therefore
\begin{eqnarray}
 c_1 \ll 1\ , \quad  
 c_2 \ll 1\ , \quad 
 c_3 \tilde{\phi} \ll 1\ , \quad 
 c_4 \tilde{\phi}^2 \ll 1 \ldots
\label{bound}
\end{eqnarray}
The first test for an inflationary model is whether these conditions
are {\em naturally} satisfied. Many complicated models have been
proposed which purport to do so \cite{modelrev}, although this is not
always evident on closer examination. We consider here the simplest
possibility employing a {\em single} inflaton field in minimal
supergravity.

\section{Natural Supergravity Inflation}

In supersymmetric theories with a single supersymmetry generator
($N=1$), complex scalar fields are the lowest components, $\phi^a$, of
chiral superfields $\Phi^a$ which contain chiral fermions, $\psi^a$,
as their other component. (We will take $\Phi^a$ to be left-handed
chiral superfields so that $\psi^a$ are left-handed fermions.) Masses
for fields will be generated by spontaneous symmetry breaking so that
the only fundamental mass scale is the Planck scale, $M$. (This is
aesthetically attractive and also what follows if the underlying
theory generating the effective low-energy supergravity theory follows
from the superstring.) The $N=1$ supergravity theory describing the
interaction of the chiral superfields is specified by the K\"{a}hler
potential
\begin{equation}
 G (\Phi ,\Phi^\dagger) = d (\Phi , \Phi^\dagger) + \ln |f(\Phi)|^2\ , 
\label{G}
\end{equation}
which yields the scalar potential
\begin{equation}
 V = {\rm e}^{d/M^2} \left[F^{A\dagger} (d_A^B)^{-1} F_B -
      3\frac{|f|^2}{M^2} \right]
 + {\rm D-terms}\ ,
\label{V}
\end{equation}
where 
\begin{equation}
 F_A \equiv \frac{\partial f}{\partial\Phi^A} 
     + \left(\frac{\partial d}{\partial\Phi^A}\right) \frac{f}{M^2}, \qquad
 \left(d_A^B\right)^{-1} \equiv 
  \left(\frac{\partial^2 d}{\partial\Phi^A \partial\Phi_B^\dagger}\right)^{-1}.
\end{equation}
Here the function $d$ sets the form of the kinetic energy terms of the
theory
\begin{equation}
 L_{\rm kin} = \frac{\partial^2 d}{\partial\phi_A \partial\phi^{\dagger B}}
  \partial_\mu \phi_A \partial^\mu \phi^{\dagger B} ,
\label{kin}
\end{equation}
while the superpotential $f$ determines the non-gauge interactions of
the theory. For canonical kinetic energy terms,
$d=\sum_A\phi_A^\dagger\phi^A$, the potential takes the relatively
simple form
\begin{equation}
 V = \exp \left({\sum_A \phi_A^\dagger \phi^A}\right)
      \left[\sum_B \left|\frac{\partial f}{\partial \phi_B}\right|^2 - 3
     \left|f\right|^2\right] .
\label{pot}
\end{equation}
In order for there to be a period of inflation, it is necessary for at
least one of the terms $|\case{\partial{f}}{\partial\phi_B}|$ to be
{\em non-zero}. However, these are precisely the order parameters for
supersymmetry so this corresponds to supersymmetry breaking during
inflation. While there are several possible mechanisms for such
breaking, it suffices for the purposes of this discussion to simply
assume that one of the terms has nonvanishing value $\Lambda^4$, where
$\Lambda$ denotes the supersymmetry breaking scale. Now expansion of
the exponential in eq.(\ref{pot}) shows that $c_2=1$ and $c_4=1$ in
eq.(\ref{expand}), in conflict with the requirement (\ref{bound}) for
successful inflation. We see that the problem can be traced to the
presence of the overall factor involving the exponential in the scalar
potential (\ref{pot}).

In ref.\cite{success} we suggested that in theories with moduli the
problem is easily avoided. Moduli are fields in superstring theories
which, in the absence of supersymmetry breaking, have no
potential. The moduli vevs serve to determine the fundamental
couplings of the theory and for the moduli of interest here they
appear in the superpotential only in combination with non-moduli
fields, serving to determine the latter's couplings in terms of their
vevs. We argued \cite{success} that the quadratic terms in the
potential involving the non-moduli fields such as the inflaton would
be absent for special values of these vevs and, since the resultant
potential would drive inflation, just this desired configuration would
come to dominate the final state of the universe. Subsequently we
realized \cite{paper} that it is not even necessary to invoke such an
`anthropic principle' because there is a {\em quasi-fixed point} in
the evolution of the moduli. As we discuss below, this ensures, for
initial values in the basin of attraction of the fixed point, that the
cancellation of the quadratic terms applies, ensuring that condition
(\ref{bound}) is satisfied. Now although the moduli have a flat
potential in the absence of supersymmetry breaking, once supersymmetry
is broken they may acquire a potential through the moduli dependence
of the $d$ function in the scalar potential (\ref{V}). This is
potentially disastrous because such a potential would drive the moduli
vevs away from the value needed to cancel the quadratic inflaton
term. However the kinetic term often has a larger symmetry than the
full Lagrangian; for example the canonical form has an $SU(N)$
symmetry where $N$ is the total number of chiral fields. In this case
there will be many moduli left massless even when supersymmetry is
broken because they will be (pseudo) Goldstone modes associated with
the spontaneous breaking of this symmetry. These moduli can play the
role discussed above eliminating the quadratic term in the inflaton
potential \cite{paper}.

The mechanism proposed in ref.\cite{paper} applies to a large class of
models, the only condition being that the kinetic term does indeed
have a symmetry leading to pseudo-Goldstone modes. We have given
\cite{paper} two specific examples to illustrate the idea in detail,
one for the case where the kinetic term has a larger symmetry than the
full Lagrangian, and another where the potential discussed above
follows from a symmetry of the full theory. In both cases the field
potential is of the general form
\begin{equation}
 V (|\tilde{\phi}|, \varphi) = \Lambda^4 \left(1 +
                         \beta |\tilde{\phi}|^2 \varphi +
                         \gamma |\tilde{\phi}|^3 +
                         \delta |\tilde{\phi}|^4 + \ldots \right)
\label{vf}
\end{equation}
where further terms have been added in the expansion of $V$. The cubic
term may arise from a cubic term in the superpotential \cite{success};
this is allowed if the additional ($U(1)$) symmetry of the $\phi$
field in an $R$-symmetry. (Alternatively there may be another modulus
with $U(1)$ charge such that a cubic term can appear in the kinetic
function $d$.) If the cubic term is not present, then the quartic
term, which is always allowed in the kinetic term by the $SU(2)$ and
$U(1)$ symmetries of the model, will dominate. Note that the
parameters $\beta$, $\gamma$ and $\delta$ are all naturally of order
unity. 

For successful `new' inflation, we are interested in initial
conditions which lead to $|\tilde{\phi}|$ being small but there is
nothing which constrains the initial conditions of $\varphi$. However
since the potential (\ref{vf}) has an infrared fixed point with
$\tilde{\phi}=\varphi=0$, {\em any} initial value of $\tilde{\phi}$
and $\varphi$ will be driven there if they are within the domain of
attraction, given (for positive $\beta$) by
\begin{equation}
 \varphi \geq \frac{3|\gamma|}{2\beta} \left[1 + 
  \sqrt{1 + \frac{4}{9}\left(\frac{\beta}{|\gamma|}\right)^2}\right]
  |\tilde{\phi}|\ .
\end{equation}
Thus, {\em without any fine tuning of the initial conditions} (beyond
the condition that the fields lie in this domain of attraction), the
fields are driven to fixed values and the potential becomes a
constant, driving a period of inflation. (We have chosen $\beta$ to be
positive, while $\gamma$ should be negative if it is to lead to an
inflationary potential.) Moreover this fixed point corresponds to a
point of inflection in the potential which is unstable with respect to
small perturbations. Thus inflation is naturally terminated by a new
mechanism as follows. The equations of motion for $\varphi$ and
$|\tilde{\phi}|$ are
\begin{equation}
 \ddot{\varphi} + 3 H \dot{\varphi} = - \beta |\tilde{\phi}|^2, \qquad 
 |\ddot{\tilde{\phi}}| + 3H|\dot{\tilde{\phi}}| 
  = - \beta \varphi |\tilde{\phi}| + 3 |\gamma| |\tilde{\phi}|^2 ,
\label{em}
\end{equation}
so while $\varphi$ is positive, the fields are driven to the fixed
point and inflation begins. However $\varphi$ has fluctuations of
order the Hawking temperature of the De Sitter vacuum, $T_{\rm
H}=H/2\pi$, so should it fluctuate and become negativ, the fields will
be driven away from the fixed point thus ending inflation. (For
$\beta$ negative, the reverse would be the case.) The initial
conditions for this stage are $\varphi,|\tilde{\phi}|\sim\,H$;
thereafter, as we see from eq.(\ref{em}), $|\tilde{\phi}|$ will grow
more rapidly than $\varphi$ and the cubic term in the potential will
soon dominate.

There are two distinctive features of the potential (\ref{vf}) which
ensure that, after the transition to positive $\varphi$, there will be
an inflationary period yielding density fluctuations of the magnitude
observed. The first is that this potential has a very small gradient
in the neighbourhood of the origin in field space so it generates a
long period of slow-roll inflation during which quantum fluctuations
are naturally small. The second feature is that the full potential,
including higher order terms, is governed by an overall scale,
$\Lambda$. The reason is that the potential arises from the $d$ term
of eq.(\ref {G}) which, in the absence of supersymmetry breaking,
gives rise to the kinetic term and thus does not contribute to the
potential, vanishing when derivatives are set to zero. Thus the
potential is proportional to the (fourth power of the) overall
supersymmetry breaking scale, $\Lambda$. This scale is plausibly of
${\cal O}(10^{14})$\,GeV {\em during} inflation \cite{success} and, in
conjunction with the small slope, correctly yields the required
magnitude of density fluctuations.

\section{Implications for Large-Scale Structure and CMB Anisotropy}

The inflationary period following from a potential of the form
(\ref{vf}) with no quadratic term (and $\gamma=-4$) has been closely
studied earlier \cite{success}. The field value when perturbations of
a given scale cross the Hubble radius is obtained by integrating the
equation of motion (\ref{em}) back from the end of inflation, which
occurs at $\tilde{\phi}_{\rm end}\simeq\,M/6|\gamma|$ when
$\epsilon=1$. Thus
$\tilde{\phi}_\star\simeq\,M/3{|\gamma|}[N_\star(k)+2]$ and using
eq.(\ref{Nstar}) we find a logarithmic (squared) deviation from scale
invariance for the scalar perturbations,
\begin{equation}
 \delta^2_{{\rm H}}(k) = \frac{9\gamma^2}{75\pi^2} 
  \frac{\Lambda^{4}}{M^{4}} [N_\star(k) + 2]^{4}\ .
\label{spec}
\end{equation}
This corresponds to a `tilted' spectrum, $\delta^2_{{\rm
H}}(k)\propto\,k^{n-1}$, with
\begin{equation}
 n (k) = 1 + 2\eta - 6\epsilon \simeq \frac{N_\star(k)-2}{N_\star(k)+2}\ ,
\end{equation}
i.e. $n\simeq0.92$ for $N_\star=51$ corresponding to the scales probed
by {\sl COBE} \cite{success}. We emphasize that a leading cubic term
in the potential gives the ${\em maximal}$ departure from
scale-invariance.  The slope of the potential is tiny,
$\epsilon=1/18\gamma^2(N_{\star}+2)^4\simeq7.0\times10^{-9}\gamma^{-2}$,
but its curvature is not: $\eta=-2/(N_{\star}+2)\simeq-0.038$.
Consequently, although the spectrum is tilted, the gravitational wave
background (\ref{grav}) is {\em negligible}. Furthermore the tilt
would be greater if $N_\star$ is smaller, for example if there is a
second epoch of `thermal inflation' when the scale-factor inflates by
$\sim20$ e-folds \cite{thermalinfl} so that the value of $N_\star$
appropriate to {\sl COBE} is 31 rather than 51, and $n\simeq0.88$. We
normalize the spectrum (\ref{spec}) to the CMB anisotropy using the
expression for the (ensemble-averaged) quadrupole,
\begin{equation}
 \frac{\langle{Q_{\rm rms}}\rangle^2}{T_0^2} = \frac{5 C_2}{4\pi} = 
   \frac{5}{4} \int_0^\infty \frac{{\rm d}k}{k}\  
   j_2^2\left(\frac{2k}{H_0}\right)\delta^2_{{\rm H}}(k)\ ,
\end{equation}
where $j_2$ is the second-order spherical Bessel function. According
to the {\sl COBE} data \cite{cobe,cobenorm}, $Q_{\rm
rms}\simeq\,20\pm2\,\mu$K for $n\simeq0.9$ which fixes the
inflationary scale to be
\begin{equation}
 \frac{\Lambda}{M} \simeq 2.8 \pm 0.14 \times 10^{-4}\ |\gamma|^{-1/2}\ ,
\end{equation} 
consistent with general theoretical considerations of supersymmetry
breaking \cite{success}.

The spectrum of the (dimensionless) rms mass fluctuations after matter
domination (per unit logarithmic interval of $k$) is given by
\cite{structure}
\begin{equation}
 \Delta^2 (k) \equiv \frac{k^3 P(k)}{2\pi^2} 
              = \delta^2_{{\rm H}}(k)\ T^2 (k) \left(\frac{k}{aH}\right)^4\ ,
\end{equation}
where $P(k)$ is the usual power spectrum and the `transfer function'
$T(k)$ takes into account that linear perturbations grow at different
rates depending on the relation between their wavelengths, the Jeans
length and the Hubble radius. For CDM we use \cite{structure},
\begin{equation}
 T (k) = \left[1 + \left\{a k + (b k)^{3/2} + (c k)^2
          \right\}^{\nu}\right]^{-1/\nu}
\end{equation}
with $a=6.4\Gamma^{-1}h^{-1}{\rm Mpc}$, $b=3\Gamma^{-1}h^{-1}{\rm
Mpc}$, $c=1.7\Gamma^{-1}h^{-1}{\rm Mpc}$ and $\nu=1.13$, where the
`shape parameter' is $\Gamma\simeq\Omega{h}\,{\rm e}^{-2\Omega_{N}}$
\cite{peadodd}. 

\begin{figure}
\makebox{
\rotate[r]{\epsfxsize2.2in\rotate[l]{(a)}\epsffile{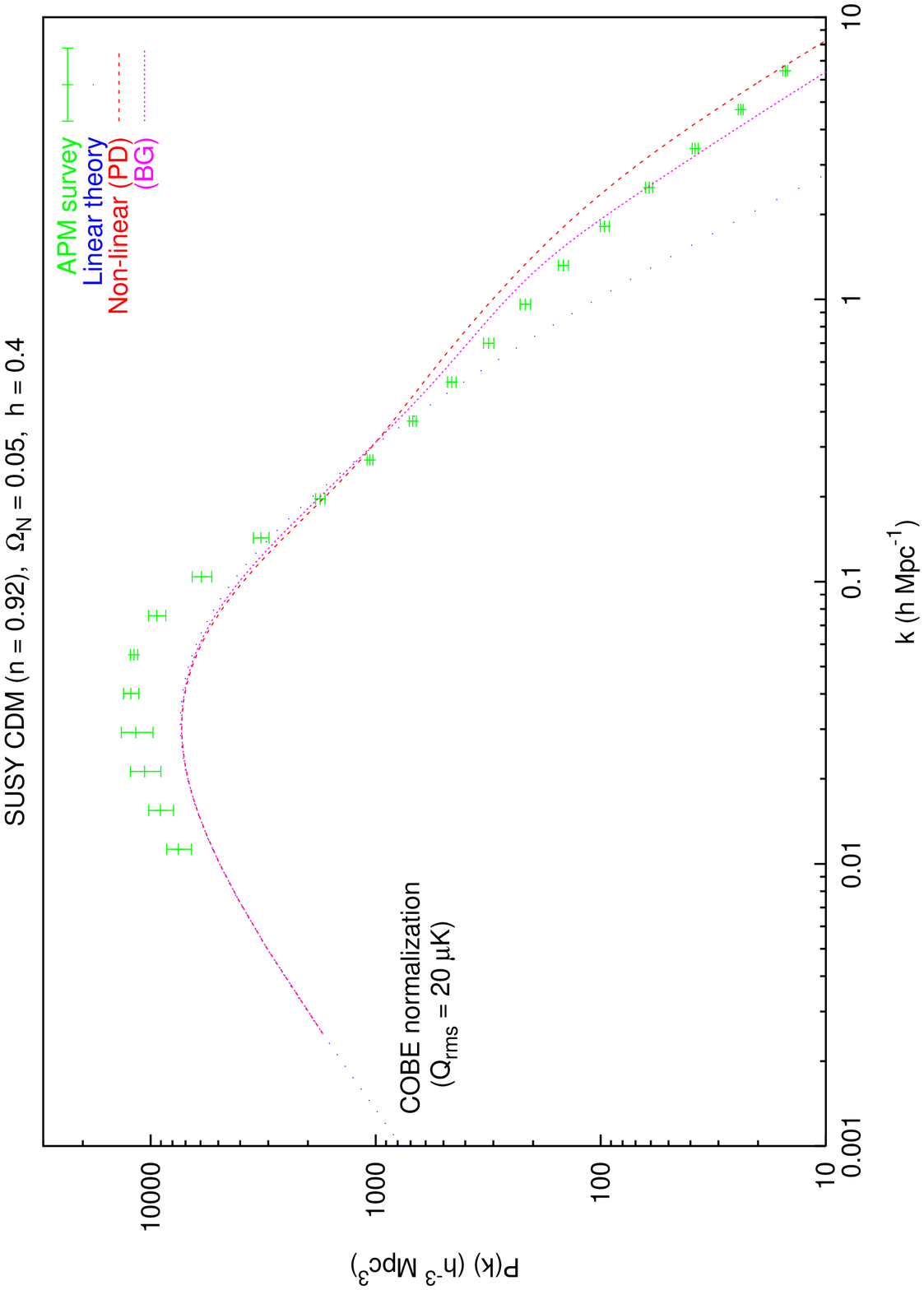}}
\rotate[r]{\epsfxsize2.2in\rotate[l]{(b)}\epsffile{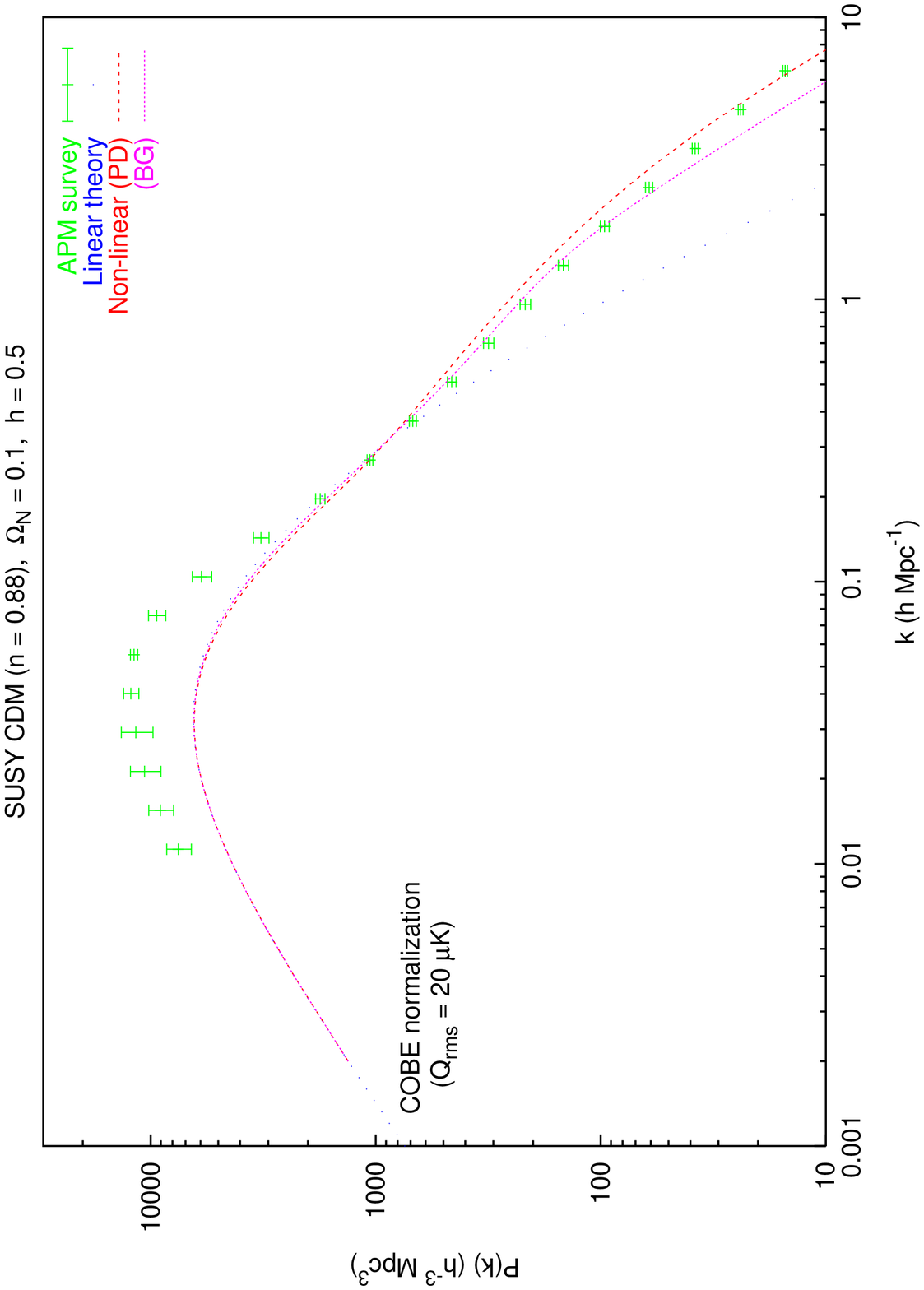}}
}
\caption{Predicted non-linear power spectra of density fluctuations in
cold dark matter, normalized to {\sl COBE} and compared with data
inferred from the APM galaxy survey.}
\label{ps}
\end{figure}

The cosmological parameters adopted for `standard' CDM are $h=0.5$ and
$\Omega_{N}=0.05$ \cite{structure}. However, observational
uncertainties still permit the Hubble parameter to be as low as 0.4
\cite{hrev}. Also the nucleon density parameter $\Omega_{\rm N}$ may
be as high as $0.033h^{-2}$, taking into account the recent upward
revision of the $^4$He mass fraction \cite{bbn}. We show $P(k)$ for
$\Omega_{N}=0.05,\,0.1$ and $h=0.4,\,0.5$ in figure~\ref{ps}, having
taken account of non-linear gravitational effects at small scales
using the prescriptions of ref.\cite{peadodd} (PD) and
ref.\cite{nonlin} (BG). The tilt in the primordial spectrum which {\em
increases} logarithmically with decreasing spatial scales allows a
good fit on scales of $\sim1-100$\,Mpc to the data points obtained
\cite{apm} from the angular correlation function of APM galaxies, if
the Hubble parameter (nucleon density) are taken to be at the lower
(upper) end of the allowed range. (Only 1$\sigma$ statistical errors
are shown; at $k\lesssim0.05h\,{\rm Mpc}^{-1}$, there are also large
systematic errors \cite{apm} so the apparent discrepancy here needs
further investigation.) Note that the expected characteristic
``shoulder'' at small scales due to the non-linear evolution is
clearly visible in the APM data. Other studies of tilted spectra
\cite{tilt,liddle} focussed on the linear evolution and/or used a
compendium \cite{peadodd} of data from different surveys (having
different systematic biases) rather than one set of high quality
data. We conclude that the problem with the excess power on small
scales in the {\sl COBE}-normalized standard CDM model \cite{white} is
naturally alleviated in supergravity inflation as anticipated earlier
\cite{success,eps}, with {\em no need for a component of hot dark
matter}.

\begin{figure}
\makebox{
\rotate[r]{\epsfxsize2.3in\rotate[l]{}\epsffile{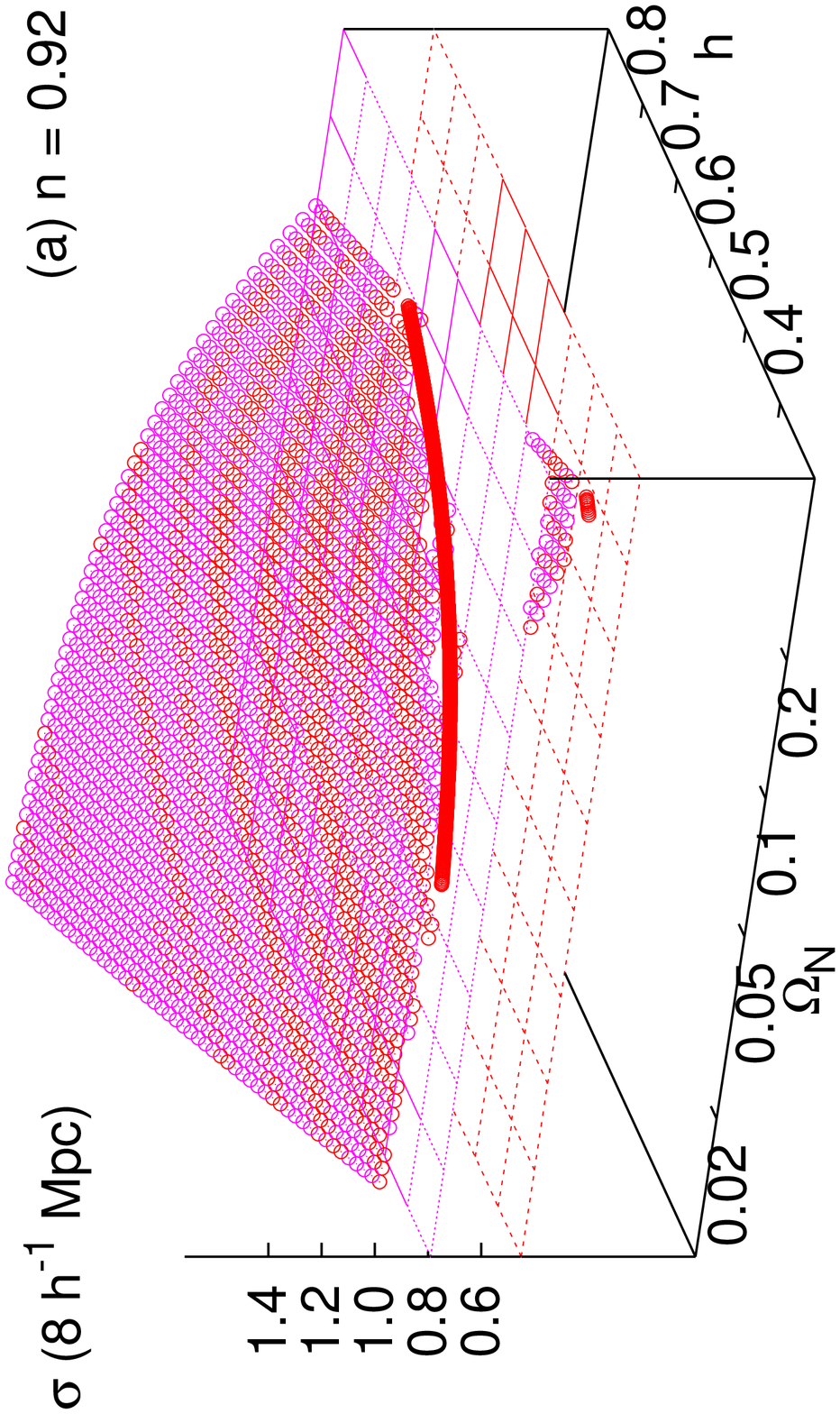}}
\rotate[r]{\epsfxsize2.3in\rotate[l]{}\epsffile{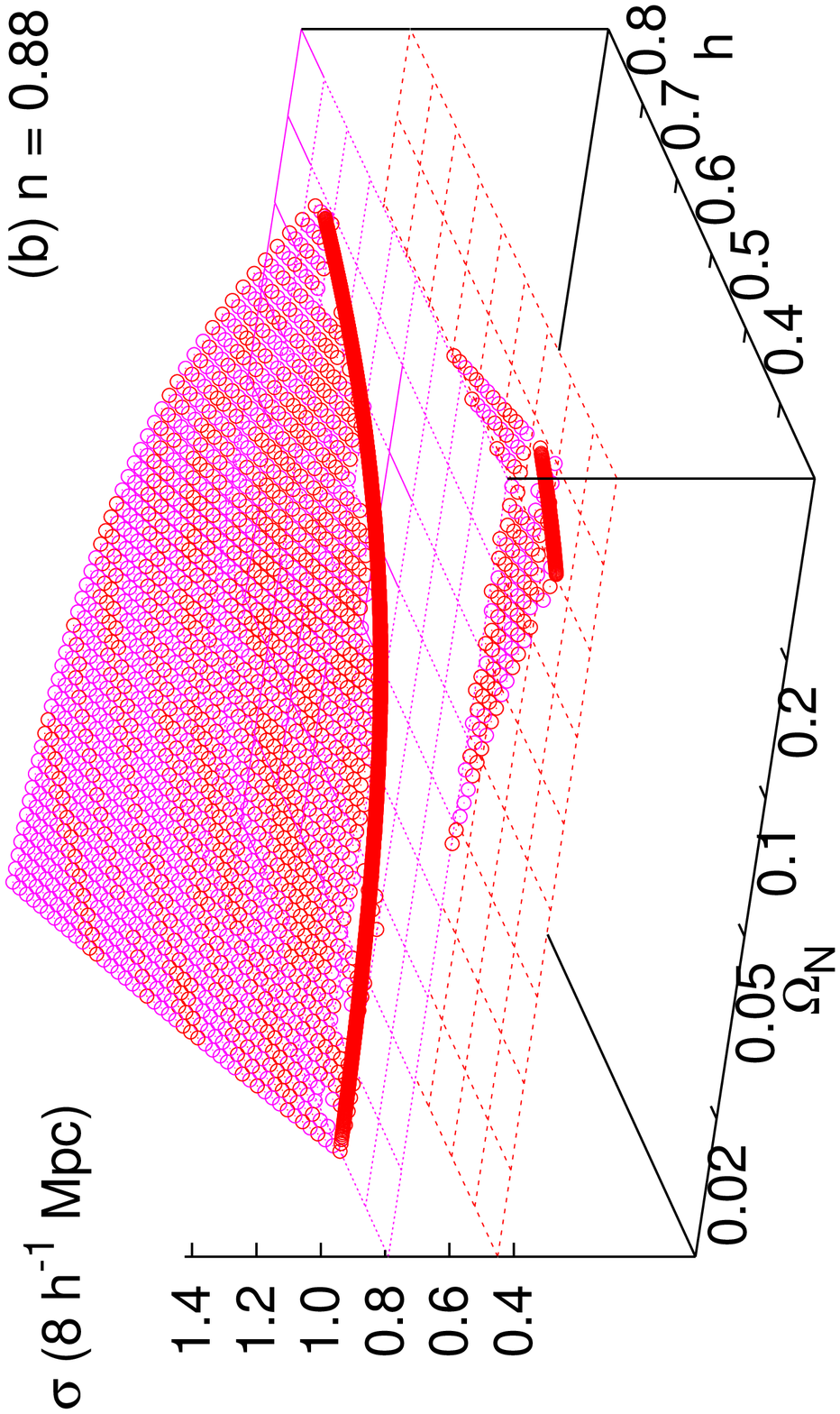}}
}
\caption{Predicted variance of the density field smoothed over a
 sphere of radius $8\,h^{-1}$\,Mpc, compared with observational limits
 (horizontal planes) inferred from rich clusters of galaxies.}
\label{sigma8}
\end{figure}

\begin{figure}
\makebox{
\rotate[r]{\epsfxsize2.2in\rotate[l]{(a)}\epsffile{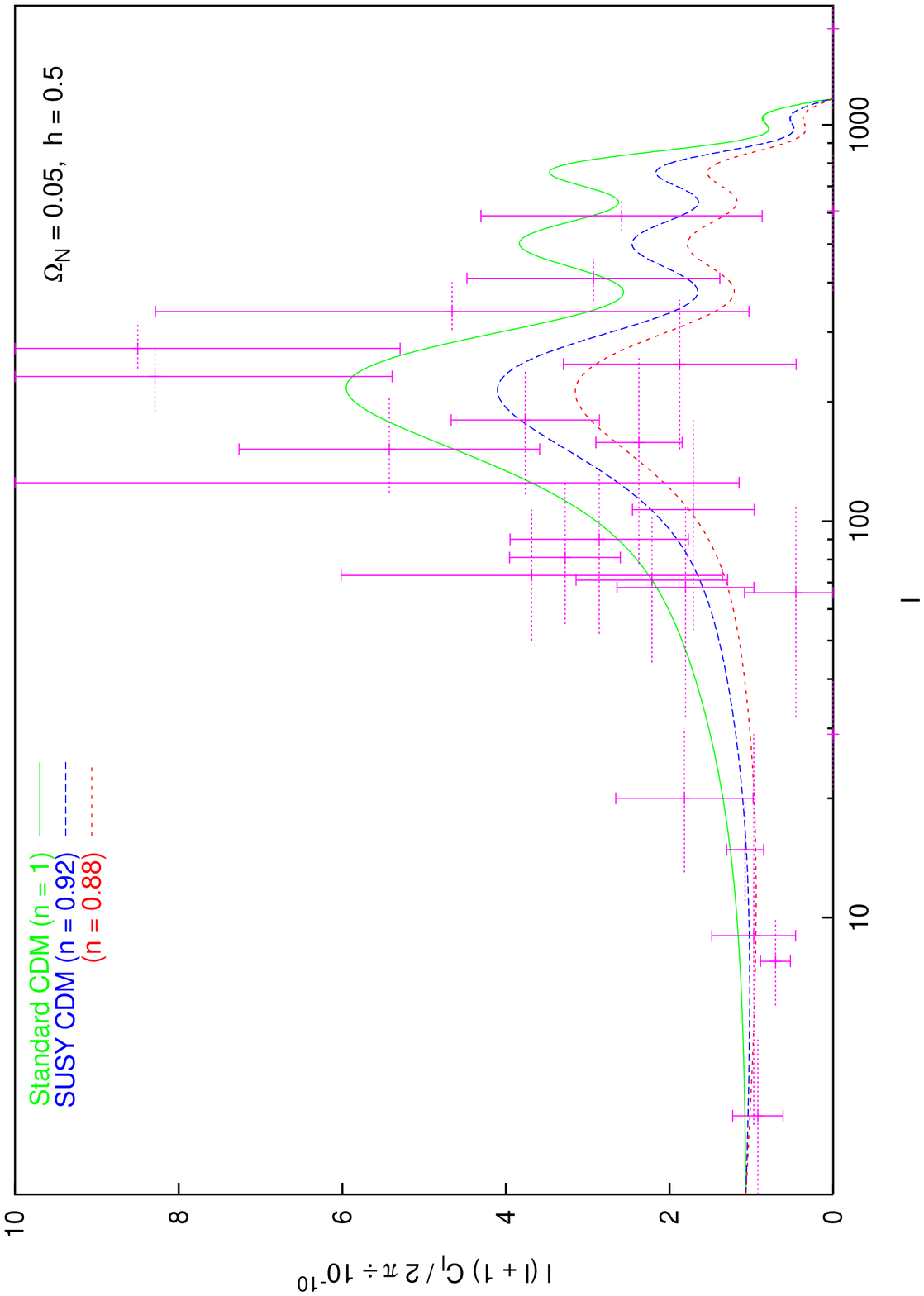}}
\rotate[r]{\epsfxsize2.2in\rotate[l]{(b)}\epsffile{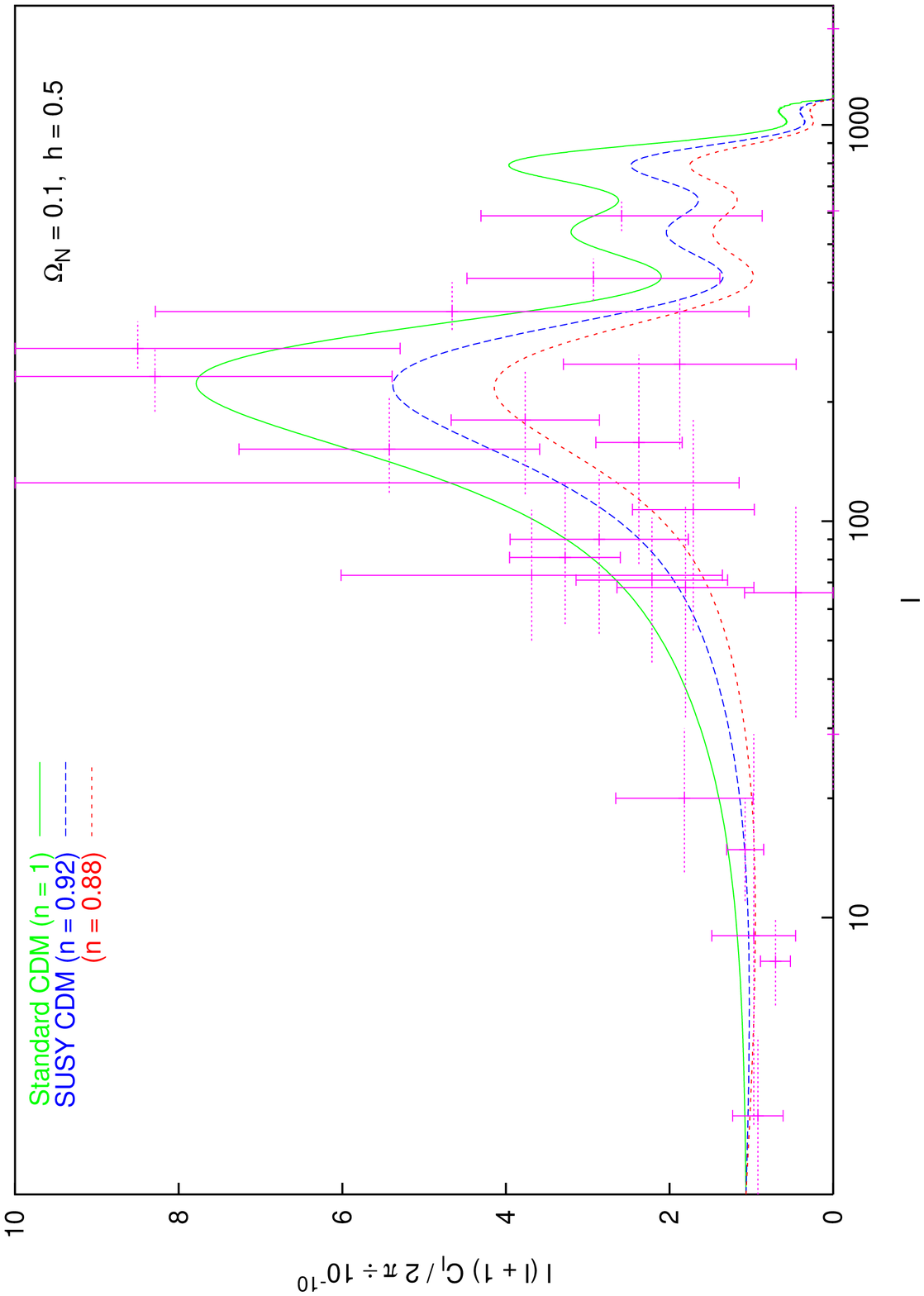}}
}
\caption{Predicted angular power spectra of CMB anisotropy,
normalized to {\sl COBE} and compared with data from current
ground-based and balloon experiments.}
\label{cmb}
\end{figure}

We also calculate some averaged quantities of observational
interest. A common measure of large-scale clustering is the variance,
$\sigma(R)$, of the density field smoothed over a sphere of radius
$R$, usually taken to be $8\,h^{-1}$\,Mpc, given in terms of the
matter density spectrum by
\begin{equation}
 \sigma^2 (R) = \frac{1}{H_0^4} \int^\infty_0 W^2 (kR)\ 
                 \delta^2_{\rm H}(k)\ T^2(k)\ k^3\ {\rm d} k\ , 
\end{equation}
where a `top hat' smoothing function,
$W(kR)=3\left[\frac{\sin(kR)}{(kR)^3}-\frac{\cos(kR)}{(kR)^2}\right]$,
has been used. As seen from figure~\ref{sigma8}, the observational
value of $\sigma\,(8 h^{-1}{\rm Mpc})=0.60^{+0.19}_{-0.15}$ ($95\%$
c.l.), inferred from the abundances of rich clusters of galaxies
\cite{white,viana} favours high tilt, high $\Omega_{\rm N}$ and low
$h$. 

Another interesting quantity is the smoothed peculiar velocity field
or `bulk flow',
\begin{equation}
 \sigma_v^2(R) = \frac{1}{H_0^2} \int^\infty_0 W^2(kR)
                 \ {\rm e}^{-(12\,h^{-1} k)^2}\ 
                 \delta^2_{\rm H}(k)\ T^2(k)\ k\ {\rm d}k\ ,
\end{equation}
where, for direct comparison with observations, we have applied an
additional gaussian smoothing on $12 h^{-1}{\rm Mpc}$. For the two
models shown in figure~\ref{ps} we find,
\begin{eqnarray}
 \sigma_v (40 h^{-1}{\rm Mpc}) &= 383 &\pm 38\ {\rm km\ sec}^{-1}\  
  (N_{\sl COBE}=51,\ \Omega_{\rm N}=0.05,\ h=0.4), \nonumber \\ 
                               &= 320 &\pm 32\ {\rm km\ sec}^{-1}\ 
  (N_{\sl COBE}=31,\ \Omega_{\rm N}=0.1,\ h=0.5).
\end{eqnarray}
to be compared with the POTENT III measurement of
$\sigma_v(40h^{-1}{\rm Mpc})=373\pm50$\,km~sec$^{-1}$
\cite{potent}. We do not consider constraints coming from the
abundances of collapsed objects at high redshift such as
Lyman-$\alpha$ clouds and quasars \cite{liddle,repeat}, as this
involves many astrophysical uncertainties at present.
 
An unambiguous test of the model is the predicted CMB anisotropy. To
compute this accurately requires numerical solution of the coupled
linearized Boltzmann, Einstein and fluid equations for the
perturbation in the photon phase space distribution. We use the
COSMICS computer codes \cite{bert} to calculate the angular power
spectrum using the primordial scalar fluctuation spectrum
(\ref{spec}). The first 1000 multipoles are plotted in
figure~\ref{cmb}, taking $\Omega_{\rm N}=0.05,\ 0.1$, along with a
compendium of recent observations \cite{cmbrev}, and the prediction of
standard CDM is shown for comparison. The height of the first `Doppler
peak' is preferentially boosted for the higher value of $\Omega_{\rm
N}$ and this is favoured by the CMB observations in conjunction with
the large-scale structure data, as has been noted independently
\cite{repeat}. For a given value of $\Omega_{\rm N}$ the effect of the
spectral tilt is to suppress the heights of all the peaks. Although
present ground-based observations are inconclusive, this prediction
will be definitively tested by the forthcoming satellite-borne
experiments, MAP and COBRAS/SAMBA.

In summary, inflationary model building has received a new impetus as
a consequence of the impressive progress in observations of
large-scale structure and CMB anisotropy which can discriminate
between such models. It appears quite plausible that within the next decade
such {\em astronomical} data will provide a direct window to physics
at the unification scale.

\acknowledgements{I thank Y.M. Cho and Jihn Kim for the invitation to
participate in this stimulating meeting, and Jennifer Adams and Graham
Ross for allowing me to present our joint work.}


\begin{references}

\bibitem{paper}
 J.A. Adams, G.G. Ross and S. Sarkar, Preprint OUTP-96-48P [hep-ph/9608336].

\bibitem{guth}
 A. Guth, Phys. Rev. D23 (1981) 347.

\bibitem{book}  
 A.D. Linde, {\sl Particle Physics and Inflationary Cosmology} 
  (Harwood Academic, 1990).

\bibitem{susyinflrev}
 For a review see, K.A. Olive, Phys. Rep. 190 (1990) 307.

\bibitem{sugrarev}
 J. Wess and J. Bagger, {\em Supersymmetry and Supergravity} (Princeton, 1993).

\bibitem{gravprob}
 S. Weinberg, Phys. Rev. Lett. 48 (1982) 1303;\\
 For a review see, S. Sarkar, Preprint OUTP-95-16P [hep-ph/9602260]. 

\bibitem{polonyiprob}
 G. Coughlan {\em et al}, Phys. Lett. B131 (1983) 59.

\bibitem{moduliprob}
 see, e.g. T. Banks, M. Berkooz and P.J. Steinhardt,
  Phys. Rev. D52 (1995) 705.

\bibitem{success}
 G.G. Ross and S. Sarkar, Nucl. Phys. B461 (1996) 597.

\bibitem{lep}
 For reviews see, W. Marciano, Annu. Rev. Nucl. Part. Sci. 41 (1991) 469;\\
 P. Langacker, M. Luo and A. K. Mann, Rev. Mod. Phys. 64 (1992) 87.

\bibitem{susyunif}
 For reviews see, S. Dimopoulos, Preprint CERN-TH-7531-94 [hep-ph/9412297];\\
 J. Ellis, Preprint CERN-TH-95-316 [hep-ph/9512335].

\bibitem{susydm}
 For reviews see, J. Ellis, Phys. Scr. T36 (1991) 14;\\
 G. Jungman, M. Kamionkowski and K. Griest, Phys. Rep. 267 (1996) 195.

\bibitem{cdm} 
 For a review see, J.P. Ostriker, Annu. Rev. Astron. Astrophys. 31 (1993) 689.

\bibitem{structure}
 T. Padmanabhan, {\sl Structure Formation in the Universe} (Cambridge, 1993).

\bibitem{cobe}
 G.F. Smoot {\it et al}, Astrophys. J. 396 (1992) L1;\\
 C.L. Bennett {\it et al} (COBE collab.), Astrophys. J. 464 (1996) L1.

\bibitem{efstathiou}
 For a review see, G.P. Efstathiou, Preprint OUAST/94/39.

\bibitem{white}
 S.D.M. White, G.P. Efstathiou and C.S. Frenk, Mon. Not. R. Astr. Soc. 
  262 (1993) 1023.

\bibitem{eps}
 S. Sarkar, 
 {\sl Proc. Intern. EPS Conf. on High Energy Physics, Brussels}, 
  ed. J. Lemonne {\em et al} (World Scientific, 1996) p.95 

\bibitem{cmbrev}
 For a review see, D. Scott and G.F. Smoot, Phys. Rev. D54 (1996) 118.

\bibitem{susybreak}
 M. Dine, W. Fischler and D. Nemechansky, Phys. Lett. 136B (1984) 169;\\
 G.D. Coughlan {\em et al}, Phys. Lett. 140B (1984) 44.

\bibitem{modelrev}
 For a review see, D.H. Lyth, Preprint LANCS-TH/9614 [hep-ph/9609431].

\bibitem{inflrev}
 For a review see, A.R. Liddle and D.H. Lyth, Phys. Rep. 231 (1993) 1.

\bibitem{cobenorm}
 E.F. Bunn and M. White, Eprint astro-ph/9607060.

\bibitem{thermalinfl}
 D.H. Lyth and E.D. Stewart, Phys. Rev. D53 (1996) 1784.

\bibitem{peadodd}
 J.A. Peacock and S.J. Dodds, Mon. Not. R. Astr. Soc. 267 (1994) 1020;
  280 (1996) L19.

\bibitem{hrev}
 For a review see, C.J. Hogan, Phys. Rev. D54 (1996) 112.

\bibitem{bbn}
 P.J. Kernan and S. Sarkar, Phys. Rev. D54 (1996) R3681.

\bibitem{nonlin}
 B. Jain, H.J. Mo and S.D.M. White, Mon. Not. R. Astr. Soc. 276 (1996) L25;\\
 C.M. Baugh and E. Gazta\~naga, Eprint astro-ph/9601085.

\bibitem{apm}
 C.M. Baugh and G.P. Efstathiou, Mon. Not. R. Astr. Soc. 265 (1993) 145.

\bibitem{tilt}
 M. White {\em et al}, Mon. Not. R. Astr. Soc. 276 (1995) L69.

\bibitem{liddle}
 A. Liddle {\em et al}, Mon. Not. R. Astr. Soc. 281 (1996) 531.

\bibitem{viana}
 P. Viana and A. Liddle, Mon. Not. R. Astr. Soc. 281 (1996) 323.

\bibitem{potent}
 A. Dekel, Annu. Rev. Astron. Astrophys. 32 (1994) 371. 

\bibitem{bert}
 E. Bertschinger, Eprint astro-ph/9506070 (http://arcturus.mit.edu/cosmics/).

\bibitem{repeat}
 M. White {\em et al}, Preprint SUSSEX-AST 96/5-2 [astro-ph/9605057].

\end{references}
\end{document}